# PRESSURE SPIKE IN THE LBNF ABSORBER CORE'S GUN DRILLED COOLING CHANNEL FROM AN ACCIDENT BEAM PULSE*

A. Deshpande†, P. Hurh, J. Hylen, A. Lee, J. Lewis, I. Rakhno, V. I. Sidorov, Z. Tang, S. Tariq
I. Tropin, FNAL, Batavia, IL 60510, USA

## Abstract

The LBNF Absorber consists of thirteen 6061-T6 aluminum core blocks. The core blocks are water cooled with de-ionized (DI) water which becomes radioactive during beam operations. The cooling water flows through gun-drilled channels in the core blocks. The cooling water is supplied by the LBNF Absorber Radioactive Water (RAW) cooling system which is designed as per ASME B31.3 Normal Fluid Service [1]. An uninhibited beam accident pulse striking the water channels was identified as a credible accident scenario. In this study, it is assumed that the beam pulse hits the Absorber directly without interacting with any of the other upstream beamline components. The beam parameters used for the LBNF beam are 120 GeV, 2.4 MW with a 1.2 s cycle time. The accident pulse lasts for 10 µs. The maximum energy is deposited in the 3rd aluminum core block. For the sake of simplicity, it is assumed that the accident pulse strikes the 1 in. ID water channel directly. The analysis here simulates the pressure rise in the water during and after the beam pulse and its effects on the aluminum piping components that deliver water to the core blocks. The weld strengths as determined by the Load and Resistance Factor Design (LRDF) and the Allowable Strength Design (ASD) are compared to the forces generated in the weld owing to the pressure spike. A transient structural analysis was used to determine the equivalent membrane, peak, and bending stresses and they were compared to allowable limits.

## OVERVIEW

The LBNF Absorber consists of 6061-T6 aluminum core blocks. The core blocks are water cooled with de-ionized (DI) water which becomes radioactive during beam operations. The cooling water flows through gun-drilled channels in the core blocks. The cooling water is supplied by the LBNF Absorber RAW cooling system which is designed as per ASME B31.3 Normal Fluid Service [1]. An uninhibited beam accident pulse striking the water channels has been identified as a credible accident scenario. In this scenario, it is assumed that the beam pulse hits the Absorber directly without interacting with any of the other upstream beamline components. This document presents analysis simulating the pressure rise in the water during and after the beam pulse and its effects on the aluminum piping components that deliver water to the core blocks.

## ANALYSIS

The beam parameters used for the LBNF beam are 120 GeV, 2.4 MW with a 1.2 s cycle time. The accident pulse lasts for 10 µs. The maximum energy is deposited in the 3rd aluminum core block. For the sake of simplicity, it is assumed that the accident pulse strikes the 1 in. ID water channel directly. This location, however, is approximately 36 cm or 14 in. from the edge of the core block. The energy deposited into the water and the aluminum at the above-described location along with other data was simulated in MARS [2]. The energy deposition data is a gaussian distribution with a peak value of 72 MJ/m$^3$ per pulse with a sigma of 20 mm. The missteered beam orientation during normal operations and accident scenario are shown in Fig 1.

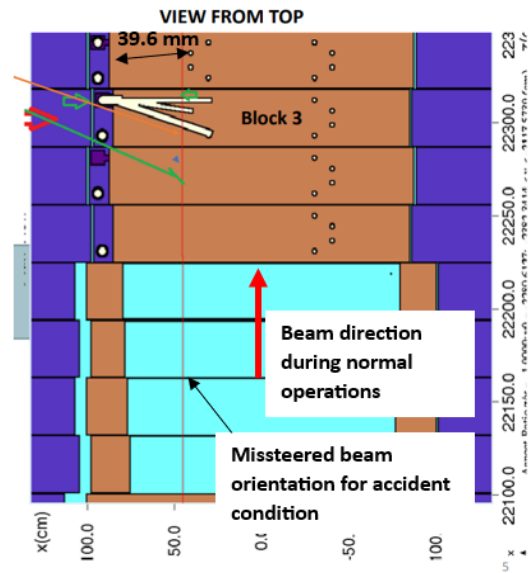

Figure 1: Top view of the Absorber core showing missteered beam during accident condition and operational beam.

### Pressure Spike in the Water Channel

A transient Computational Fluid Dynamic (CFD) analysis was done to simulate the pressure spike in the water channel when the accident pulse beam strikes it. The pressure rise in the water was tracked from the end of the beam pulse, 10 µs up to 250 µs. The inlet velocity, temperature, and outlet pressure in the channel were assumed to be 15 m/s, 25 °C, and 35 Psig, respectively. It must be noted that the actual velocity of water flowing through the channel would not exceed 3 m/s. Since the velocity of the instantaneous pressure wave front was simulated to be 1500 m/s, the initial velocity of 15 m/s in the water channel has a negligible effect on the pressure spike.

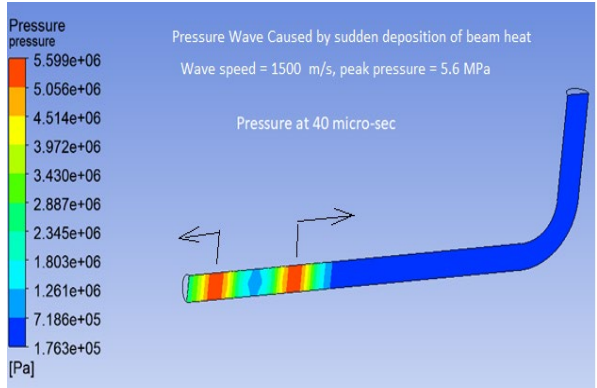

Figure 2: Pressure spike in cooling line at 40 μs caused by sudden deposition of energy.

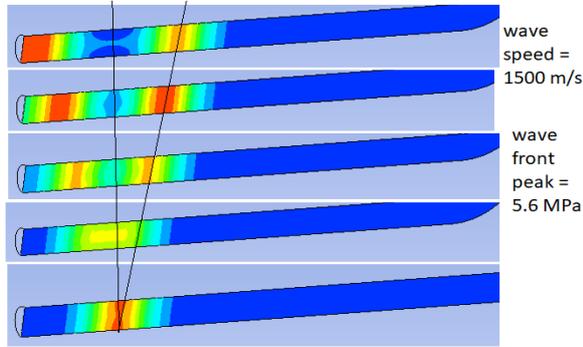

Figure 3: Wave speed and wavefront peak.

The maximum wavefront pressure simulated was 5.6 MPa or 812 Psig. Figure 2 and Fig 3 highlight this. This is the pressure realized at the 1 in. elbow. However, in the actual geometry, several 1 in. gun-drilled lines converge into a 2 in. opening. This expansion would attenuate the pressure value. Thus, the 812 Psig value is assumed to be the worst-case pressure rise.

## Effect of Pressure Rise on Aluminum Components

The effect of the above pressure spike is determined on the elbow-core block assembly, the body of the elbow, and the Sch40 aluminum pipe connected to the elbow. The elbow is welded to the core block as shown in Fig 4:

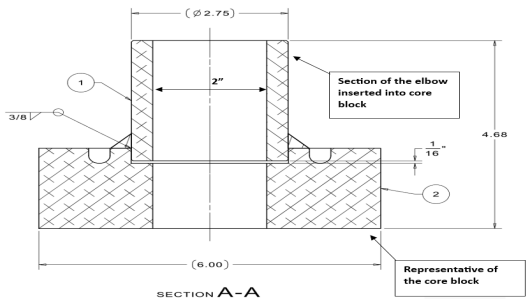

Figure 4: A representative drawing of the elbow welded to the side of the core block.

Aluminum Design Manual [3] is used to determine the strength of the weld and the base metal using Load and Resistance Factor Design (LRFD) and Allowable Strength Design (ASD) methods. The weld strength value is also determined from ASME BPVC VIII, Div I [4]. The minimum of these five values is picked as the strength of the joint assembly. This minimum strength value is compared to the forces generated by applying the 812 Psig pressure on the pipe. It must be noted that the governing code for the LBNF Absorber RAW cooling system is ASME B31.3 Normal Fluid Service [1]. The weld strengths per different criteria are shown in Table 1. Selecting the lowest value from Table 1, 8979 lb., one can determine the allowable stress in the fillet weld by dividing it by the weld area obtained from Eq. (2), 2.29 in$^2$. This value is 3921 Psi. A steady state structural analysis is performed to determine the stresses in the weld joint and in the elbow subjected to a load of 812 Psi in addition to other constraints.

$$R_{nb} = 0.6F_{tuw}S_wL_{we} \quad (1)$$

$$S_wL_{we} = 0.707a\pi D \quad (2)$$

$$R_{nw} = 0.51F_{tuw}S_wL_{we} \quad (3)$$

$$R_{nb}[LRFD] = \emptyset R_{nb} \quad (4)$$

$$R_{nb}[ASD] = \frac{R_{nb}}{\Omega} \quad (5)$$

$$R_{nw}[LRFD] = \emptyset R_{nw} \quad (6)$$

$$R_{nw}[ASD] = \frac{R_{nw}}{\Omega} \quad (7)$$

$$F_{MAWL} = 0.49S_wL_{we}S \quad (8)$$

Table 1: Calculated Weld Strengths

| Parameter | Value |
|---|---|
| Base metal strength per LRFD | 24737 lb |
| Base metal strength per ASD | 16914 lb |
| Weld metal strength per LRFD | 21026 lb |
| Weld metal strength per ASD | 14377 lb |
| Fillet weld strength per BPVC VIII, Div I | 8979 lb |

A steady state structural analysis is performed to determine the stresses in the weld joint and in the elbow subjected to a load of 812 Psi in addition to other constraints. Figure 5 highlights von Mises stresses in the elbow body and the welded joint.

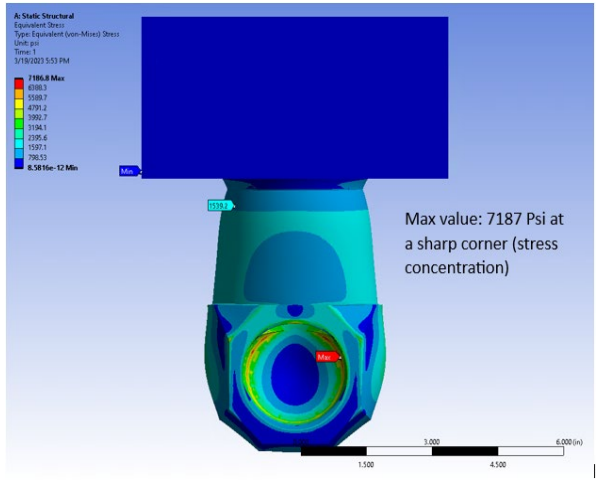

Figure 5: von Mises stresses in assembly. Stress concentration and stress in the weld.

The above analysis shows that the maximum stress developed in the fillet weld, 1540 Psi, is lower than the allowable stress of 3921 Psi. Also, the maximum stress concentration in the elbow body or 7187 Psi is lower than the allowable stress, 8000 Psi, in 6061-T6 material as per Table A-1 of ASME B31.3 [1]. One may also compare this value to 21,000 Psi, which is the allowable limit for 6061-T6 as per ASME BPVC II D [5].

A transient structural analysis is performed on the geometry shown in Fig 5. The pressure input is applied to all internal surfaces. This pressure profile is highlighted in Fig 6 below. It must be noted that the maximum value occurs at a sharp edge. As per ASME BPVC VIII, Div II [6] design by analysis section, the summation of local primary membrane plus bending must not exceed 1.5 times the tabulated allowable stress for material listed in ASME BPVC II D [5] for materials that have a ratio of yield to ultimate tensile stress greater than 0.70. And as per ASME BPVC II D [5], Table 2B, the value of allowable stress, S, for 6061-T6 material is 14,000 Psi. Thus, the allowable limit is 21,000 Psi.

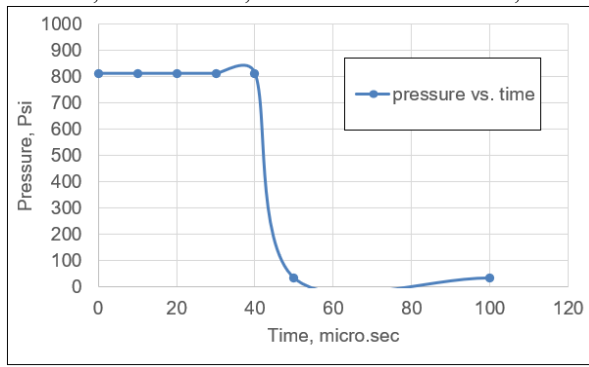

Figure 6: Pressure vs. time profile selected.

To determine the summation of membrane and bending stresses, a Stress Classification Line (SCL) is drawn in the region with the highest stress peak. The summation of the membrane and bending stress along the SCL are highlighted in Fig 7. The summation of membrane and bending is less than the allowable limit of 21,000 Psi.

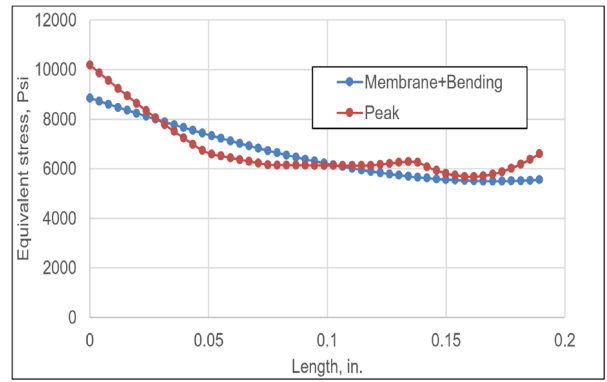

Figure 7: Sum of equivalent membrane and bending stresses, and peak stresses along the SCL.

## CONCLUSION

Pressure spike in the water channel due to accidental beam pulse can be categorized as a dynamic/impact effect under Para 301.5.1 of ASME B31.3 [1]. Para 301.5.1 only mandates that dynamic impacts shall be considered. Appendix F of ASME B31.3 [1] offers guidance and cautionary statements but does not mandate how to take them into account. Thus, the above analysis shows that stresses developed in the elbow, fillet weld connecting elbow to the block, and the pipe are below allowable limits. In addition, the summation of the membrane and bending stress generated from a transient structural analysis was found to be below the allowable limits for 6061-T6 aluminum.

## ACKNOWLEDGEMENTS

MARS modeling, energy deposition results, and ideas to determine the appropriate mesh sizing to capture the pressure spike were provided by Igor Rakhno, Nikolai Mokhov, and Igor Tropin of the Modeling/Energy Deposition/Theory group.

## REFERENCES

[1] American Society of Mechanical Engineers B31.3 Process Piping an American National Standard, 2022, American Society of Mechanical Engineers, New York, NY 10016-5990.

[2] N. V. Mokhov and S. I. Striganov, "MARS15 Overview," in *AIP Conference Proceedings*, 2007. doi:10.1063/1.2720456

[3] Aluminum Design Manual, 2015, Arlington, VA 22209.

[4] American Society of Mechanical Engineers Boiler and Pressure Vessel Code VIII, Div I., an American National Standard, 2021, American Society of Mechanical Engineers, New York, NY 10016-5990.

[5] American Society of Mechanical Engineers Boiler and Pressure Vessel Code VIII, Div II. Part D., an American National Standard, 2021, American Society of Mechanical Engineers, New York, NY 10016-5990.

[6] American Society of Mechanical Engineers Boiler and Pressure Vessel Code VIII, Div II., an American National Standard, 2021, American Society of Mechanical Engineers, New York, NY 10016-5990.